\documentclass[11pt,twocolumn]{article}
\usepackage{amssymb}
\title{\LARGE \bf Associated special functions and coherent states}
\author{\large NICOLAE COTFAS\\
\large Faculty of Physics\\ 
\large University of Bucharest\\
\large PO Box 76-54, Postal Office 76, Bucharest\\ 
\large Romania\\
\large ncotfas@yahoo.com\quad http://fpcm5.fizica.unibuc.ro/\,{\Huge ${}_{{}_{\tilde{}}}$}\,ncotfas\\[7mm]
\normalsize {\it Abstract:} \  \ A hypergeometric type equation satisfying certain conditions
defines either a finite or\\ \normalsize an infinite system of orthogonal polynomials. \ \     
We present in a unified and explicit way all these\\ \normalsize  systems of orthogonal 
polynomials, the associated special functions
and some systems of coherent\\ \normalsize   states. \ 
 This general formalism allows us to extend  
some results known only in particular cases.
\hspace {0in} \mbox{}\\[5mm]
\normalsize lowering operators, Creation and annihilation operators,
Hypergeometric-type equations.\quad \qquad \ \   \mbox{}} 
\date{}
\begin{document}
\maketitle

{\Large \bf 
\noindent 1 \ Introduction}\\
\noindent Many problems in quantum mechanics and
mathematical physics lead to equations of the type
\begin{equation}\label{hypeq}
\sigma (s)y''(s)+\tau (s)y'(s)+\lambda y(s)=0 
\end{equation}
where $\sigma (s)$ and $\tau (s)$ are polynomials of at most second
and first degree, respectively, and $\lambda $ is a constant. 
These equations are usually called {\em equations of hypergeometric
type} [15], and each can be reduced to the self-adjoint form 
\begin{equation}
[\sigma (s)\varrho (s)y'(s)]'+\lambda \varrho (s)y(s)=0 
\end{equation}
by choosing a function $\varrho $ such that 
\begin{equation}
 [\sigma (s)\varrho (s)]'=\tau (s)\varrho (s).
\end{equation}

The equation (\ref{hypeq}) is usually considered on an interval $(a,b)$,
chosen such that 
\begin{equation}\begin{array}{r}
\sigma (s)>0\qquad {\rm for\ all}\quad s\in (a,b)\\
\varrho (s)>0\qquad {\rm for\ all}\quad s\in (a,b)\\
\lim_{s\rightarrow a}\sigma (s)\varrho (s)
=\lim_{s\rightarrow b}\sigma (s)\varrho (s)=0.
\end{array}
\end{equation}
Since the form of the equation (\ref{hypeq}) is invariant under a 
change of variable $s\mapsto cs+d$, it is sufficient to analyse the cases
presented in table 1.
Some restrictions must be imposed on $\alpha $, $\beta $ in
order for the interval $(a,b)$ to exist.
We prove that
equation (\ref{hypeq}) defines an infinite sequence of orthogonal polynomials
\[ \Phi _0,\ \ \Phi _1,\ \ \Phi _2,\ ... \]
in the case $\sigma (s)\in \{ 1,\ s,\ 1-s^2\}$, and a finite one 
\[ \Phi _0,\ \ \Phi _1,\ \ ...,\ \ \Phi _L \]
in the case $\sigma (s)\in \{ s^2-1,\ s^2,\ s^2+1\}$.

The literature discussing special function theory and its application to mathematical
and theoretical physics is vast, and there are a multitude of different conventions
concerning the definition of functions. A unified approach 
is not possible without a unified definition for the associated special functions.
In this paper we define them as 
\begin{equation}
 \Phi _{l,m}(s)=\left(\sqrt{\sigma (s)}\right)^m\, \frac{{\rm d}^m}{{\rm d}s^m}\Phi _l(s)
\end{equation}
where  $\Phi _l$ are the orthogonal polynomials defined by equation  (\ref{hypeq}). 
The table 1 allows 
one to pass in each case from our parameters $\alpha $, $\beta $ to the parameters
used in different approach.

\noindent {\it Table 1:} Particular cases 
{\footnotesize (in each case  $\tau (s)=\alpha s\!+\!\beta $).}
\begin{table}[h]
\begin{center}
{\footnotesize
\begin{tabular}{|c|c|c|c|}
\hline \hline
$\begin{array}{l}
\mbox{}\\[-2mm]
\sigma (s)\\[-2mm]
\mbox{}
\end{array}$ & $\varrho (s)$ & 
$\alpha ,\beta $ &  $(a,b) $\\
\hline \hline
$\begin{array}{l}
\mbox{}\\[-1mm]
1\\[-3mm]
\mbox{}
\end{array}$ 
& ${\rm e}^{\frac{1}{2}\alpha s^2+\beta s}$ & $\alpha <0$
& $\mathbb{R}$\\ 
\hline 
$\begin{array}{l}
\mbox{}\\[-3mm]
s\\[-3mm]
\mbox{}
\end{array}$ 
& $s^{\beta -1} {\rm e}^{\alpha s}$ & 
$\begin{array}{l}
\alpha \!<\!0\\
\beta \!>\!0
\end{array}$& $(0,\infty )$\\ 
\hline
$\begin{array}{l}
\mbox{}\\[-3mm]
1\!\!-\!\!s^2\\[-3mm]
\mbox{}
\end{array}$ 
& $\begin{array}{l}
\mbox{}\\[-2mm]
(1\!+\!s)^{-\frac{\alpha -\beta }{2}-1}\times \\[1mm]
\times (1\!-\!s)^{-\frac{\alpha +\beta }{2}-1}\\[-2mm]
\mbox{}
\end{array}$ & 
$\begin{array}{l}
\alpha \!<\! \beta \\
\alpha \!+\!\beta \!<\!0
\end{array}$ & $(-1,1)$\\ 
\hline  
$\begin{array}{l}
\mbox{}\\[-1mm]
s^2\!\!\!-\!\!1\\[-3mm]
\mbox{}
\end{array}$ 
& $\begin{array}{l}
\mbox{}\\[-2mm]
(s\!+\!1)^{\frac{\alpha -\beta }{2}-1}\times \\[1mm]
\times (s\!-\!1)^{\frac{\alpha +\beta }{2}-1}\\[-2mm]
\mbox{}
\end{array}$ &
$\begin{array}{l}
0\!<\! \alpha \!+\!\beta \\
\alpha <0
\end{array}$ & $(1,\infty )$\\ 
\hline 
$\begin{array}{l}
\mbox{}\\[-3mm]
s^2\\[-3mm]
\mbox{}
\end{array}$ 
& $s^{\alpha -2}{\rm e}^{-\frac{\beta }{s}}$ & $
\begin{array}{l}
\alpha \!<\!0\\ 
\beta \!>\!0
\end{array}$ & $(0,\infty )$\\
\hline 
$\begin{array}{l}
\mbox{}\\[-1mm]
s^2\!\!\!+\!\!1\\[-3mm]
\mbox{}
\end{array}$ 
& $\begin{array}{l}
\mbox{}\\[-2mm]
(1+s^2)^{\frac{1}{2}\alpha -1}\times \\[1mm]
\mbox{}\quad \times {\rm e}^{\beta \arctan s}\\[-2mm]
\mbox{}
\end{array}$ & 
$\alpha <0$ & $\mathbb{R}$\\
\hline
\end{tabular} }
\end{center}
\label{table}
\end{table}

\noindent
In our previous papers [6,~7], we presented a systematic study of the Schr\"odinger
equations exactly solvable in terms of associated special functions following
Lorente [14], Jafarizadeh and Fakhri [12]. In the present paper, our aim is to extend 
this unified formalism by including a larger class of creation/annihilation operators
and some temporally stable coherent states of Gazeau-Klauder type [1,~8--10,~13].\\[1cm]
%
{\Large \bf 
2 \ Orthogonal polynomials\\[1mm]
\mbox{}\ \ \ \ of hypergeometric-type}\\
\noindent 
Let $\tau (s)=\alpha s+\beta $ be a fixed polynomial, and let
\begin{equation}
\lambda _l\!=-\frac{\sigma ''(s)}{2}l(l-1)-\tau '(s)l
\!=-\frac{\sigma ''}{2}l(l-1)-\alpha \,l
\end{equation}
for any $l\in \mathbb{N}$. It is well-known [15] that for $\lambda =\lambda _l$,
the equation (\ref{hypeq}) admits a polynomial solution 
$\Phi _l=\Phi _l^{(\alpha ,\beta )}$ of at most $l$ degree
\begin{equation} \label{eq3}
\sigma (s) \Phi _l ''+\tau (s) \Phi _l '+\lambda _l\Phi _l=0.
\end{equation}
If the degree of the polynomial $\Phi _l$ is $l$ then it satisfies the
Rodrigues formula [15]
\begin{equation}
\Phi _l(s)=\frac{B_l}{\varrho (s)}\frac{{\rm d}^l}{{\rm d}s^l}[\sigma ^l(s)\varrho (s)]
\end{equation}
where $B_l$ is a constant. Based on the relation 
\[ \{ \ \delta \in \mathbb{R}\ |\ 
\lim_{s\rightarrow a}\sigma (s)\varrho (s)s^\delta  
=\lim_{s\rightarrow b}\sigma (s)\varrho (s)s^\delta =0 \ \}\]
\[ =\left\{
\begin{array}{lll}
[0,\infty ) & {\rm if} & \sigma (s)\in \{ 1,\ s,\ 1-s^2\}\\[2mm] 
[0,-\alpha ) & {\rm if} & \sigma (s)\in \{ s^2-1,\ s^2,\ s^2+1\}
\end{array} \right.  \]
one can prove [7] that the system of polynomials $\{\Phi _l\ |\ l<\Lambda \}$, where
\begin{equation}
\Lambda \!=\!\left\{ \begin{array}{lcl}
\infty & {\rm for} & \sigma (s)\in \{ 1,\ s,\ 1-s^2\}\\[2mm]
\frac{1-\alpha }{2} & { \rm for } & 
\sigma (s)\in \{ s^2\!-\!1,\ s^2,\ s^2\!+\!1\} 
\end{array}\right.
\end{equation}
is orthogonal with weight function $\varrho (s)$ in $(a,b)$. This means that
equation (\ref{hypeq}) defines an infinite sequence of orthogonal polynomials
\[ \Phi _0,\ \ \Phi _1,\ \ \Phi _2,\ ... \]
in the case $\sigma (s)\in \{ 1,\ s,\ 1-s^2\}$, and a finite one 
\[ \Phi _0,\ \ \Phi _1,\ \ ...,\ \ \Phi _L \]
with \ $L=\max \{ l\in \mathbb{N}\ |\ l<(1-\alpha )/2\}$
in the case $\sigma (s)\in \{ s^2-1,\ s^2,\ s^2+1\}$.

The polynomials $\Phi _l^{(\alpha ,\beta )}$ can be expressed in terms of the 
classical orthogonal polynomials as
{\small 
\begin{equation}\label{classical}
\begin{array}{l}
  \Phi _l^{(\alpha ,\beta )}(s)=\\[3mm]
\mbox{}\qquad \left\{ \begin{array}{lll}
H_l\left(\sqrt{\frac{-\alpha }{2}}\, s-\frac{\beta }{\sqrt{-2\alpha }}\right)  
& {\rm if} & \sigma (s)=1\\[2mm]
L_l^{\beta -1}(-\alpha s)  & {\rm if} & \sigma (s)=s\\[2mm]
P_l^{(-\frac{\alpha +\beta }{2}-1,\ \frac{-\alpha +\beta }{2}-1)}(s)  & {\rm if} & \sigma (s)=1\!\!-\!s^2\\[2mm]
P_l^{(\frac{\alpha -\beta }{2}-1,\ \frac{\alpha +\beta }{2}-1)}(-s)  & {\rm if} & \sigma (s)=s^2\!\!-\!1\\[2mm]
\left(\frac{s}{\beta }\right)^lL_l^{1-\alpha -2l}\left(\frac{\beta }{s}\right) 
& {\rm if} & \sigma (s)=s^2\\[2mm]
{\rm i}^lP_l^{(\frac{\alpha +{\rm i}\beta }{2}-1,\ \frac{\alpha -{\rm i}\beta }{2}-1)}({\rm i}s) 
& {\rm if} & \sigma (s)=s^2\!\!+\!1
\end{array} \right.
\end{array}
\end{equation}
}
where $H_n$, $L_n^p $ and $P_n^{(p,q)}$ are the Hermite,
Laguerre and Jacobi polynomials, respectively.\\[1cm]
%
{\Large \bf 
3 \ Associated special functions,\\[1mm]
\mbox{}\ \ \ \ raising and lowering operators}\\
Let $l\in \mathbb{N}$, $l<\Lambda $, and let $m\in \{ 0,1,...,l\}$.
The functions
\begin{equation}\label{def}
\Phi _{l,m}(s)=\kappa ^m(s)\frac{{\rm d}^m}{{\rm d}s^m}\Phi _l(s) 
\end{equation}
where
\[  \kappa (s)=\sqrt{\sigma (s)}\]  
are called the {\em associated special functions}. 
If we  differentiate (\ref{eq3}) $m$ times and then multiply 
the obtained relation by $\kappa ^m(s)$ then we get the equation
\begin{equation}\label{Hm}
H_m \Phi _{l,m}=\lambda _l\Phi _{l,m}
\end{equation}
where $H_m$ is the differential operator
\[
H_m =-\sigma (s) \frac{d^2}{ds^2}-\tau (s) \frac{d}{ds}
+\frac{m(m-2)}{4}\frac{(\sigma '(s))^2}{\sigma (s)} \ \ \ \ \ \mbox{}\]   
\begin{equation}
\label{defHm}
 + \frac{m\tau (s)}{2}\frac{\sigma '(s)}{\sigma (s)}
-\frac{1}{2}m(m-2)\sigma ''(s)-m\tau '(s) .
\end{equation}

The relation 
\begin{equation} \label{scalarprod}
 \langle f,g\rangle 
=\int_a^b\overline{f(s)}\, g(s)\varrho(s)ds
\end{equation}
defines a scalar product on the space 
\[ \mathcal{H}_m={\rm span}\{ \Phi _{l,m} \ |\ m\leq l <\Lambda \} \]
spanned by $\{ \Phi _{l,m} \ |\ m\leq l <\Lambda \}.$ 
For each $m<\Lambda $, the special functions $\Phi _{l,m}$ with $m\leq l<\Lambda $ 
are orthogonal with weight function $\varrho (s)$ in $(a,b)$, and
the functions corresponding to consecutive values of $m$ are related 
through the raising/lowering operators [6,~7,~12]
\begin{equation}\begin{array}{l}
A_m=\kappa (s)\frac{d}{ds}-m\kappa '(s)\\[3mm]
A_m^+=-\kappa (s)\frac{d}{ds}-\frac{\tau (s)}{\kappa (s)}-(m-1)\kappa '(s)
\end{array}
\end{equation}
namely, 
\begin{equation}\label{AmAm+}
\begin{array}{l}
A_m\Phi _{l,m}=\left\{ \begin{array}{lll}
0 & {\rm for} & l=m\\
\Phi _{l,m+1} & {\rm for} & m<l<\Lambda 
\end{array} \right. \\[5mm]
A_m^+\Phi _{l,m+1}\!=\!(\lambda _l\!-\!\lambda _m)\Phi _{l,m}\ \  
{\rm for}\ \ 0\leq m<l< \Lambda .
\end{array}
\end{equation}
In addition, we have the relations [5,~11]
\begin{equation}\label{philm}
\Phi _{l,m}=
\frac{A_m^+ }{\lambda _l-\lambda _m}
\frac{A_{m+1}^+ }{\lambda _l-\lambda _{m+1}}...
\frac{A_{l-1}^+ }{\lambda _l-\lambda _{l-1}}\Phi _{l,l}
\end{equation}
for \ $0\leq m<l< \Lambda $, and
\begin{equation}\label{fact}
H_m-\lambda _m=A_m^+A_m\qquad H_{m+1}-\lambda _m
=A_mA_m^+ 
\end{equation} 
\begin{equation}\label{interw}
H_mA_m^+=A_m^+H_{m+1}\qquad A_mH_m
=H_{m+1}A_m
\end{equation} 
for $m+1<\Lambda .$ 

The functions 
\begin{equation}
 \phi _{l,m}=\Phi _{l,m}/||\Phi _{l,m}||
\end{equation}
where 
\begin{equation}
 ||f||=\sqrt{\langle f,f\rangle }
\end{equation}
are the {\em normalized associated special functions}. 
Since [6,~7]
\begin{equation} \label{norm}
||\Phi _{l,m+1}||
=\sqrt{\lambda _l-\lambda _m}\, ||\Phi _{l,m}||\qquad  
\end{equation}
they satisfy the relations
\begin{equation}\label{relations} \begin{array}{l}
A_m\phi _{l,m}\!=\!\left\{ \begin{array}{lll}
0 & {\rm for} & l=m\\
\sqrt{\lambda _l\!-\!\lambda _m}\phi _{l,m+1} & {\rm for} & m<l<\Lambda 
\end{array} \right.\\[3mm]
A_m^+\phi _{l,m+1}\!=\!\sqrt{\lambda _l\!-\!\lambda _m}\phi _{l,m}
\quad {\rm for}\  0\leq m<l<\Lambda \\[3mm]
\phi _{l,m}=
\frac{A_m^+ }{\sqrt{\lambda _l-\lambda _m}}
\frac{A_{m+1}^+ }{\sqrt{\lambda _l-\lambda _{m+1}}}...
\frac{A_{l-1}^+ }{\sqrt{\lambda _l-\lambda _{l-1}}}\phi _{l,l}. 
\end{array}
\end{equation}
\mbox{}\\[5mm]
%
{\Large \bf 
4 \ Coherent states in the\\[1mm]
\mbox{}\ \ \ \ case $\sigma (s)\in \{ 1,\, s,\, 1-s^2\}$}\\
Let $m$ be a fixed natural number and $\gamma $ a fixed real number.
The sequence 
\[ \phi _{m,m},\quad \phi _{m+1,m},\quad \phi _{m+2,m},\ \dots \]
is a complete orthonormal sequence in the Hilbert space 
\[ \mathcal{H}\!=\!\left\{ \varphi :(a,b)\longrightarrow \mathbb{C}\ \left| 
\ \int_a^b|\varphi (s)|^2\varrho (s)\, ds<\infty \right. \right\} \]
with scalar product (\ref{scalarprod}), for any $m\in \mathbb{N}$.\\[5mm]
%
\noindent{\large \bf 4.1 \ Creation and annihilation\\
\mbox{}\qquad  operators}\\
The linear operators (see figure 1)
\begin{equation}
 \begin{array}{l}
a_m,\ a_m^+:\mathcal{H}_m\longrightarrow  \mathcal{H}_m\\[3mm]
\mbox{}\qquad  a_m=U_m^{-1}A_m\qquad a_m^+=A_m^+U_m 
\end{array} 
\end{equation}
defined by using the unitary operator
\begin{equation}
 \begin{array}{l}
U_m:\mathcal{H}_m\longrightarrow  \mathcal{H}_{m+1}\\[3mm]
\mbox{}\qquad  U_m\phi _{l,m}={\rm e}^{-{\rm i}\gamma (\lambda _{l+1}-\lambda _l)}\phi _{l+1,m+1}
\end{array} 
\end{equation}
are mutually adjoint, 
\begin{equation} \begin{array}{l}\label{repsu11}
a_m\phi _{l,m}\!\!=\!\!\left\{ \begin{array}{ll}
0 & {\rm for} \ l\!=\!m\\
\sqrt{\lambda _{l}\!-\!\!\lambda _m} {\rm e}^{{\rm i}\gamma (\lambda _l-\lambda _{l-1})} \phi _{l-1,m} &
{\rm for} \ l\!>\!m
\end{array} \right.\\[5mm]
a_m^+\phi _{l,m}\!\!=\!\!\sqrt{\lambda _{l+1}\!-\!\!\lambda _m} 
{\rm e}^{-{\rm i}\gamma (\lambda _{l+1}-\lambda _l)} \phi _{l+1,m}\   
{\rm for}  \  l\!\geq \!m
\end{array}
\end{equation}
and
\begin{equation}
 H_m-\lambda _m=a_m^+a_m \qquad \qquad \ \mbox{}
\end{equation} 
\begin{equation}
\begin{array}{l}
[a_m^+,a_m]\phi _{l,m}=(\lambda _l-\lambda _{l+1})\phi _{l,m}\\[2mm]
\mbox{}\qquad \qquad \quad \ =(\sigma ''l+\alpha )\phi _{l,m}. 
\end{array}
\end{equation}
Since the operator $R_m=[a_m^+,a_m]$ satisfies the relations
\begin{equation}
 [R_m,a_m^+]=\sigma ''a_m^+\qquad [R_m,a_m]=-\sigma ''a_m 
\end{equation}
the Lie algebra $\mathbb{L}_m$ generated by $\{ a_m^+,a_m\}$
is finite dimensional.\\[3mm]

\begin{figure}
\setlength{\unitlength}{1mm}
\begin{picture}(90,55)(12,0)
\put(18.7,7){$a_0 $}
\put(11.8,7){$a_0^+$}
\put(28,18.3){$A_0^+$}
\put(28,12){$A_0$}
\put(24.9,8){$U_0$}
\put(18.7,22){$a_0 $}
\put(11.8,22){$a_0^+$}
\put(28,33.3){$A_0^+$}
\put(28,27){$A_0$}
\put(24.9,23){$U_0$}
\put(18.7,37){$a_0 $}
\put(11.8,37){$a_0^+$}
\put(28,48.3){$A_0^+$}
\put(28,42){$A_0$}
\put(24.9,38){$U_0$}
\put(43.7,22){$a_1 $}
\put(36.8,22){$a_1^+$}
\put(53,33.3){$A_1^+$}
\put(53,27){$A_1$}
\put(49.9,23){$U_1$}
\put(43.7,37){$a_1 $}
\put(36.8,37){$a_1^+$}
\put(53,48.3){$A_1^+$}
\put(53,42){$A_1$}
\put(49.9,38){$U_1$}
\put(68.7,37){$a_2 $}
\put(61.8,37){$a_2^+$}
\put(78,48.3){$A_2^+$}
\put(78,42){$A_2$}
\put(74.9,38){$U_2$}
\put(16,52){$.$}
\put(16,51){$.$}
\put(16,50){$.$}
\put(41,52){$.$}
\put(41,51){$.$}
\put(41,50){$.$}
\put(66,52){$.$}
\put(66,51){$.$}
\put(66,50){$.$}
\put(91,52){$.$}
\put(91,51){$.$}
\put(91,50){$.$}
\put(14.5,0){$\phi _{0,0}$}
\put(14.5,15){$\phi _{1,0}$}
\put(14.5,30){$\phi _{2,0}$}
\put(14.5,45){$\phi _{3,0}$}
\put(39.5,15){$\phi _{1,1}$}
\put(39.5,30){$\phi _{2,1}$}
\put(39.5,45){$\phi _{3,1}$}
\put(64.5,30){$\phi _{2,2}$}
\put(64.5,45){$\phi _{3,2}$}
\put(89.5,45){$\phi _{3,3}$}
\put(16,4){\vector(0,1){9}}
\put(18,13){\vector(0,-1){9}}
\put(22,15){\vector(1,0){17}}
\put(39,17){\vector(-1,0){17}}
\put(22,3){\vector(3,2){16}}
\put(16,19){\vector(0,1){9}}
\put(18,28){\vector(0,-1){9}}
\put(22,30){\vector(1,0){17}}
\put(39,32){\vector(-1,0){17}}
\put(22,18){\vector(3,2){16}}
\put(16,34){\vector(0,1){9}}
\put(18,43){\vector(0,-1){9}}
\put(22,45){\vector(1,0){17}}
\put(39,47){\vector(-1,0){17}}
\put(22,33){\vector(3,2){16}}
\put(41,19){\vector(0,1){9}}
\put(43,28){\vector(0,-1){9}}
\put(47,30){\vector(1,0){17}}
\put(64,32){\vector(-1,0){17}}
\put(47,18){\vector(3,2){16}}
\put(41,34){\vector(0,1){9}}
\put(43,43){\vector(0,-1){9}}
\put(47,45){\vector(1,0){17}}
\put(64,47){\vector(-1,0){17}}
\put(47,33){\vector(3,2){16}}
\put(66,34){\vector(0,1){9}}
\put(68,43){\vector(0,-1){9}}
\put(72,45){\vector(1,0){17}}
\put(89,47){\vector(-1,0){17}}
\put(72,33){\vector(3,2){16}}
\end{picture}
\caption{The operators $A_m$, $A_m^+$, $a_m$, $a_m^+$ and $U_m$ relating the 
functions $\phi _{l,m}$.}
\label{figure}
\end{figure}
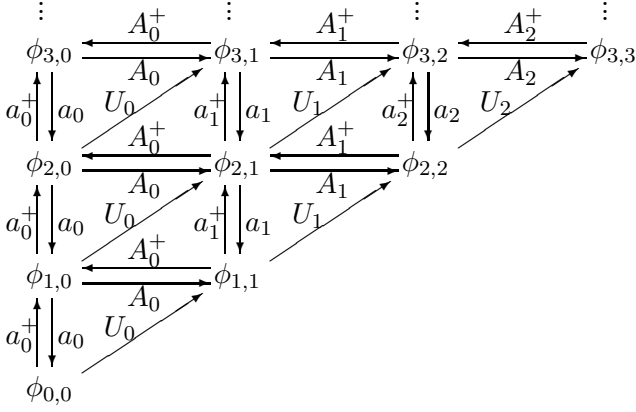

\noindent {\bf Theorem 1.} 
{\it The Lie algebra $\mathbb{L}_m$ is isomorphic} 
\[ 
\mbox{}\ \ to \ \ \left\{ \begin{array}{lcl}
h(2) & if & \sigma (s)\!\in \!\{ 1,\, s\}\\[2mm]
su(1,1) & if & \sigma (s)\!=\!1\!-\!s^2.
\end{array} \right.
\]
\mbox{}\\[3mm]  
{\em Proof}.
In the case $\sigma (s)\in \{ 1,s\}$ the operator $R_m$ is a constant
operator, namely, $R_m=\alpha $. Since $\alpha <0$, the operators 
$P_+=\sqrt{-1/\alpha }\, a_m^+$, $P_-=\sqrt{-1/\alpha }\, a_m$ and  
the unit operator $I$ form a basis of $\mathbb{L}_m$ such that
\[ [P_+,P_-]=-I\qquad [I,P_\pm ]=0 \]
that is, $\mathbb{L}_m$ is isomorphic to the Heisenberg-Weyl algebra 
$h(2)$. If $\sigma (s)=1-s^2$ then  
$K_+=a_m^+$,  $K_-=a_m $ and $K_0=R_m$ form a basis of $\mathbb{L}_m$
such that
\[ [K_+,K_-]=-2 K_0 \qquad [K_0,K_\pm ]=\pm K_\pm . \qquad \quad \rule{3pt}{3pt}\]

The operator $a_m$ can be regarded as an annihilation operator, 
and $a_m^+$ as a creation operator.\\[8mm]
%
\noindent{\large \bf 4.2 \ Coherent states}\\
Let $m\in \mathbb{N}$ be a fixed natural number. 
The functions \ $|0\rangle $, $|1\rangle $, $|2\rangle $, $\cdots $ , where 
\begin{equation}
|n\rangle =\phi _{m+n,m}
\end{equation}
satisfy the relations
\begin{equation}
\begin{array}{l}
a_m|n\rangle =\left\{ \begin{array}{ll}
0 & {\rm if} \ \ n=0\\
\sqrt{e_n}\,{\rm e}^{{\rm i}\gamma (e_n-e_{n-1})} |n-1\rangle & {\rm if} \ \ n>0
\end{array} \right. \nonumber \\[5mm]
a_m^+|n\rangle = \sqrt{e_{n+1}}\, {\rm e}^{-{\rm i}\gamma (e_{n+1}-e_n)} |n+1\rangle \\[2mm]
(H_m-\lambda _m)|n\rangle =e_n|n\rangle \nonumber 
\end{array}
\end{equation}
where
\begin{equation}\begin{array}{l}
e_n=\lambda _{m+n}-\lambda _m\\
\mbox{}\quad =\left\{ 
\begin{array}{lcl}
-\alpha n & {\rm if} & \sigma(s)\in \{ 1,s\}\\[2mm]
n(n+2m-\alpha -1) & {\rm if } & \sigma (s)=1-s^2 .
\end{array} \right.
\end{array} 
\end{equation}

For each $z\in \mathbb{C}$, the function 
\begin{equation}
 |z, \gamma \rangle=\sum_{n=0}^\infty \frac{z^n\, 
{\rm e}^{-{\rm i}\gamma e_n}}{\sqrt{\varepsilon _n}}|n\rangle 
\end{equation}
with
\begin{equation}
 \varepsilon _n=\left\{ \begin{array}{lll}
1 & {\rm if} & n=0\\
e_1e_2...e_n & {\rm if } & n>0\, .
\end{array} \right. 
\end{equation}
is an eigenfunction of $a_m$
\begin{equation}
a_m|z, \gamma \rangle =z|z, \gamma \rangle 
\end{equation}
and 
\begin{equation}
\langle z, \gamma |z, \gamma \rangle 
\!=\!\left\{ \begin{array}{ll}
{\rm e}^{-\frac{|z|^2}{\alpha }}& {\rm if} \  \sigma (s)\!\in \!\{ 1,\, s\}\\[3mm]
\mbox{}_0F_1(2m\!-\!\alpha ;|z|^2) & {\rm if} \  \sigma (s)\!=\!1\!-\!s^2.
\end{array} \right.
\end{equation} 
where 
\begin{equation}
\mbox{}_0F_1(c;\!z)\!\!=\!\!1\!+\frac{1}{c}\frac{z}{1!}+\frac{1}{c(\!c\!+\!\!1\!)}\frac{z^2}{2!}
+\frac{1}{c(\!c\!+\!\!1\!)(\!c\!+\!\!2\!)}\frac{z^3}{3!}+\!\cdots 
\end{equation} 
is the confluent hypergeometric function.\\
By using the notation $z=r{\rm e}^{{\rm i}\theta }$ and 
the modified Bessel function
\begin{equation} 
K_\nu (z)=\frac{\pi }{2}\frac{I_{-\nu }(z)-I_\nu (z)}{{\rm sin}\, (\nu \pi )}
\end{equation}
where
\begin{equation}
I_\nu (z)=\sum_{n=0}^\infty \frac{\left(\frac{1}{2}z\right)^{\nu +2n}}
{n!\, \Gamma (\nu +n+1)}
\end{equation}
we prove (following [2]) that 
$\{ \ |z, \gamma \rangle \ |\ z\in \mathbb{C}\}$
is a system of coherent states.\\[1mm]

\noindent {\bf Theorem 2.} {\it 
The system of functions
\[ \{ \ |z, \gamma \rangle \ |\ z\in \mathbb{C}\}\]
satisfies the resolution of identity
\begin{equation}
\int_\mathbb{C} d\mu \, |z, \gamma \rangle \langle z, \gamma |=I
\end{equation}
for $d\mu $ defined as
\begin{equation}\label{last}
d\mu =\!\left\{ \begin{array}{l}
\frac{-1}{\pi \alpha }{\rm e}^{\frac{1}{\alpha }|z|^2}d({\rm Re}\, z)\, d({\rm Im}\, z )\\[2mm]
\mbox{}\qquad \qquad  \quad for \ \sigma (s)\!\in \!\{ 1, s\}\\[3mm]
\frac{2r^{2m-\alpha }}{\pi \Gamma (2m-\alpha )}
K_{\frac{\alpha +1}{2}-m}(2r)\, dr\, d\theta \\[2mm]
\mbox{}\qquad \qquad  \quad for \  \sigma (s)\!=\!1\!-\!s^2.
\end{array} \right.
\end{equation} 
}
\mbox{}\\[3mm]
{\em Proof.} If $\sigma (s)\in \{ 1,s\}$ then 
\begin{equation}\label{cs1}
|z, \gamma \rangle =\sum_{n=0}^\infty
\frac{z^n\, {\rm e}^{-{\rm i}\gamma e_n}}{\sqrt{n!\, (-\alpha )^n}}|n\rangle .
\end{equation}
By denoting $t=-\frac{r^2}{\alpha }$ and using the integration by parts
we get
\[ \begin{array}{l} 
\frac{-1}{\pi \alpha }\!\int_\mathbb{C} \!d({\rm Re}\, z)\, 
d({\rm Im}\, z)|z,\!\gamma \rangle \langle z,\!\gamma |\\[5mm]
\mbox{}\  =\frac{-1}{\pi \alpha }\!\!\sum_{n,n'} {\rm e}^{-{\rm i}\gamma (e_n-e_{n'})}\times \\[5mm]
\times \!\!\left(\int_0^\infty \!\!{\rm e}^\frac{r^2}{\alpha }\!\frac{r^{n+n'+1}}
{\sqrt{n!\, n'!\, (-\alpha )^{n+n'}}}dr \!\!
\int_0^{2\pi }\!\!{\rm e}^{{\rm i}(n-n')\theta }
d\theta \! \right)\!|n\rangle \langle n'|\\[5mm]
\mbox{}\  =\frac{-2}{\alpha }\sum_n
\left( \int_0^\infty \!\!\!{\rm e}^\frac{r^2}{\alpha }\frac{1}{n!}
\left(\frac{r^2}{-\alpha }\right)^n\!\!r\, dr\!\right) |n\rangle \langle n|\\[5mm]
\mbox{}\  =\sum_n\left(\int_0^\infty \!\!\!\! {\rm e}^{-t}\frac{t^n}{n!}\, 
dt \right)|n\rangle \langle n| 
\!=\!\sum_n |n\rangle \langle n|=I.
\end{array} \]
If $\sigma (s)=1-s^2$ then 
\begin{equation}\label{cs2}\begin{array}{l}
|z, \gamma \rangle =\sqrt{\Gamma (2m-\alpha )}
\sum_{n=0}^\infty \frac{z^n\, {\rm e}^{-{\rm i}\gamma e_n}}{\sqrt{n!\, \Gamma (n+2m-\alpha )}}|n\rangle 
\end{array}
\end{equation}
Denoting $d\mu =\mu (r)\, dr\, d\theta $ we get 
\[ \begin{array}{l}
\int_\mathbb{C} d\mu \, |z, \gamma \rangle \langle z, \gamma |\\[5mm]
=\sum_{n=0}^\infty \frac{2\pi\Gamma (2m-\alpha )}
{n!\, \Gamma (n+2m-\alpha )}
\left(\int_0^\infty r^{2n}\mu (r)\, dr \right) |n\rangle \langle n| 
\end{array} \]
and hence, we must have the relation (Mellin transformation)
\begin{equation} \label{mellin}
2\pi\Gamma (2m-\alpha )\!\!
\int_0^\infty \!\!\!r^{2n}\mu (r)\, dr \!=\! \Gamma (n+1)\, \Gamma (n+2m-\alpha ). 
\end{equation}
The formula [4]
\[ \int_0^\infty \!\!\!2x^{\eta +\xi }K_{\eta -\xi }(2\sqrt{x})\, x^{n-1}dx\!=\!
\Gamma (2\eta +n)\, \Gamma (2\xi +n) \]
for $x=r^2$, \ $\eta =\frac{1}{2}$, \ $\xi =m-\frac{\alpha }{2}$ becomes
{\small 
\begin{equation}\label{mellin1}
4\!\!\!\int_0^\infty \!\!\!\!\!\!\!r^{2n}K_{\frac{\alpha +1}{2}-m}(2r)\, r^{2m-\alpha } dr
\!=\!\Gamma (n+1)\Gamma (n+2m-\alpha ).
\end{equation}}
The relations (\ref{mellin}) and (\ref{mellin1}) lead to (\ref{last}).\qquad \quad $\rule{3pt}{3pt}$\\[5mm]

If we consider the `number' operator [1,~8]
\begin{equation}
N:\mathcal{H}\longrightarrow \mathcal{H}\qquad N|n\rangle =n|n\rangle 
\end{equation}
that is,
\begin{equation}
 N=\sum_{n=0}^\infty n\, |n\rangle \langle n| 
\end{equation}
then the operator $H=H_m-\lambda _m$ can be written as 
\[ \begin{array}{l}
H=\sum_{n=0}^\infty e_n|n\rangle \langle n|\\[5mm] 
\mbox{}\ \  =\!\left\{ \begin{array}{lll}
-\alpha N & {\rm if} & \sigma (s)\in \{ 1,\, s\}\\[3mm]
N(N+2m-\alpha -1) & {\rm if} & \sigma (s)=1-s^2.
\end{array} \right. 
\end{array}\]

The operators $a_m$ and $a_m^{\perp }$, where [8]
\[ a_m^{\perp }\!=\!\frac{N}{H}a_m^+ 
\!=\!\left\{ \begin{array}{lll}
-\frac{1}{\alpha }a_m^+ & {\rm if} & \sigma (s)\!\in \!\{ 1,\, s\}\\[3mm]
\frac{1}{N+2m-\alpha -1}a_m^+ & {\rm if} & \sigma (s)\!=\!1\!-\!s^2.
\end{array} \right. \]
satisfy the relations 
\[ [a_m,a_m^{\perp }]\!=\!I\qquad [N,a_m^{\perp }]\!=\!a_m^{\perp }\qquad  [N,a_m]\!=\!-a_m. \]
Therefore, we can consider the non-unitary displacement operator [8]
\[  \begin{array}{l}
D(z)={\rm exp}(z\, a_m^{\perp }-\overline{z}\, a_m)\\[5mm]
\mbox{}\qquad ={\rm exp}\left( -\frac{1}{2}|z|^2\right)\ 
{\rm exp}(z\, a_m^{\perp })\ {\rm exp}(-\overline{z}\, a_m) 
\end{array}\]
and 
\begin{equation}
|z, \gamma \rangle =D(z)\, |0\rangle \qquad {\rm for\ any\ } z\in \mathbb{C}. 
\end{equation}

Since the Hermitian operators 
\begin{equation}
 X=\frac{1}{\sqrt{2}}(a_m^++a_m) \qquad P=\frac{\rm i}{\sqrt{2}}(a_m^+-a_m) 
\end{equation}
satisfy the commutation relation
\begin{equation}
[X,P]={\rm i}[a_m,a_m^+] 
\end{equation}
and $|z, \gamma \rangle $ are eigenstates of $a_m$,
the coherent states $|z, \gamma \rangle $  minimize the uncertainty relation [8]
\begin{equation}
 (\Delta X)^2(\Delta P)^2\geq \frac{1}{4}\langle {\rm i}[X,P]\rangle ^2.
\end{equation}

The presence of the phase factor in definition of $|z, \gamma \rangle $ leads to
the temporal stability of these coherent states 
\begin{equation}
 {\rm e}^{-{\rm i}tH}|z, \gamma \rangle =|z, \gamma +t\rangle .
\end{equation}
\mbox{}\\[2mm]
\noindent{\large \bf 4.3 \ Analytical representations}\\
The space 
\[ \mathcal{F}_m\!=\!\left\{ f:\mathbb{C}\longrightarrow \mathbb{C} \left|
\begin{array}{l}
f \ {\rm is \ an \ analytic \ function}\\ 
\int_\mathbb{C}|f(z)|^2\, d\mu <\infty 
\end{array}
\right. \right\}
\]
is a Hilbert space with the inner product 
\begin{equation}
(f,g)=\int_\mathbb{C}\overline{f(z)}\, g(z)\, d\mu 
\end{equation}
where $d\mu $ is the measure defined by (\ref{last}).\\ 
Following Bargmann [3], we associate to each\\ $\varphi :(a,b)\longrightarrow \mathbb{C}$ from  $\mathcal{H}$ 
\begin{equation}
 \varphi =\sum_{n=0}^\infty c_n|n\rangle \qquad {\rm with}\qquad  
\sum_{n=0}^\infty |c_n|^2<\infty 
\end{equation}
the entire function  $f:\mathbb{C}\longrightarrow \mathbb{C}$
\begin{equation}
 f(z)\!=\!\langle \overline{z}, \gamma |\varphi \rangle 
\!=\!\!\sum_{n=0}^\infty \frac{z^n{\rm e}^{{\rm i}\gamma e_n}}{\sqrt{\varepsilon _n}}\langle n|\varphi \rangle
\!=\!\!\sum_{n=0}^\infty \frac{z^n{\rm e}^{{\rm i}\gamma e_n}}{\sqrt{\varepsilon _n}}c_n .
\end{equation}
Particularly, the function 
\begin{equation}
u_n:\mathbb{C}\longrightarrow \mathbb{C}\qquad u_n(z)=\frac{z^n{\rm e}^{{\rm i}\gamma e_n}}{\sqrt{\varepsilon _n}}
\end{equation}
corresponds to $\varphi =|n\rangle $, for any $n\in \mathbb{N}$.

From the relation (\ref{mellin}) we get
\begin{equation}
(z^n,z^k)=\varepsilon _n\, \delta _{nk}=\left\{
\begin{array}{ll}
\varepsilon _n & {\rm if}\ n=k\\
0 & {\rm if}\ n\not=k
\end{array}\right. 
\end{equation}
whence 
\begin{equation}
(f,g)=\sum_{n=0}^\infty \varepsilon _n\, \overline{c_n}\, d_n
\end{equation}
for any two elements $f,g\in \mathcal{F}_m$
\[ f=\sum_{n=0}^\infty c_nz^n\quad  {\rm and}\quad  
g=\sum_{n=0}^\infty d_nz^n . \] 
For an entire function $f=\sum_{n=0}^\infty c_nz^n$ we have
\begin{equation}
f\in \mathcal{F}_m \quad \Longleftrightarrow \quad \sum_{n=0}^\infty \varepsilon _n\, |c_n|^2<\infty .
\end{equation}
Since
\begin{equation}
(u_n,f)=\sqrt{\varepsilon _n}\,{\rm e}^{-{\rm i}\gamma e_n} c_n
\end{equation}
the orthonormal system $u_0,\, u_1,\, u_2,\, ...$ is complete 
\begin{equation}
||f||^2=\sum_{n=0}^\infty \varepsilon _n\, |c_n|^2=\sum_{n=0}^\infty |(u_n,f)|^2.
\end{equation}

The isometry 
\[ \mathcal{H}\longrightarrow \mathcal{F}_m:\ \ 
\sum_{n=0}^\infty c_n|n\rangle \mapsto \sum_{n=0}^\infty c_n\, u_n \]
allows us to identify the two Hilbert spaces, and to get
\begin{equation}
\begin{array}{l}
a_m\, u_n =\left\{ \begin{array}{ll}
0 & {\rm if} \ \ n=0\\
\sqrt{e_n}\,{\rm e}^{{\rm i}\gamma (e_n-e_{n-1})} u_{n-1} & {\rm if} \ \ n>0
\end{array} \right. \nonumber \\[5mm]
a_m^+\, u_n = \sqrt{e_{n+1}}\, {\rm e}^{-{\rm i}\gamma (e_{n+1}-e_n)} u_{n+1} \\[2mm]
R_m\, u_n=(e_n-e_{n+1})u_n
\end{array}
\end{equation}
whence
\begin{equation}\begin{array}{l}
a_m\, z^n=e_n\, z^{n-1}\\
a_m^+\, z^n=z^{n+1}\\
R_m\, z^n=(e_n-e_{n+1})z^n .
\end{array}
\end{equation}
From the relation 
\begin{equation}
e_n=\left\{ 
\begin{array}{lcl}
-\alpha n & {\rm if} & \sigma(s)\in \{ 1,s\}\\[2mm]
n(n+2m-\alpha -1) & {\rm if } & \sigma (s)=1-s^2 .
\end{array} \right.
\end{equation}
one gets
\begin{equation}
a_m=-\alpha \frac{d}{dz}\qquad a_m^+=z\qquad R_m=\alpha  
\end{equation}
in the case $\sigma(s)\in \{ 1,s\}$, and 
\begin{equation}\begin{array}{l}
a_m=z\frac{d^2}{dz^2}+(2m-\alpha )\frac{d}{dz}\\[2mm]
a_m^+=z\\[2mm]
R_m=-2z\frac{d}{dz}-2m+\alpha 
\end{array}
\end{equation}
in the case $\sigma (s)=1-s^2$.\\[1cm]
%
{\Large \bf 5 \ Coherent states in the\\[1mm]
\mbox{}\ \ \ \ case $\sigma (s)\in \{ s^2-1,\, s^2,\, s^2+1\}$}\\
Let $\tau (s)=\alpha s+\beta $ be a fixed polynomial,
\begin{equation}
L=\max \{ l\in \mathbb{N}\ |\ l<(1-\alpha )/2\}
\end{equation}
and let $m$ be a fixed natural number with $0\leq m\leq L.$
The functions $|0\rangle $, $|1\rangle $, ..., $|\mathcal{L}\rangle $, where 
$\mathcal{L}=L-m$ and 
\begin{equation}
|n\rangle =\phi _{m+n,m}
\end{equation}
are orthogonal and span a $(\mathcal{L}+1)$-dimensional space 
\begin{equation}
\mathcal{E}_m={\rm span}\{ |0\rangle , |1\rangle , ..., |\mathcal{L}\rangle \}. 
\end{equation}
\mbox{}\\[3mm]
\noindent{\large \bf 5.1 \ Creation and annihilation\\
\mbox{}\qquad operators}\\
By following [16] and the analogy with the case $\sigma (s)\in \{ 1,\, s,\, 1-s^2\}$
we define the creation and annihilation operators
$\tilde{a}_m^+,\, \tilde{a}_m:\mathcal{E}_m\longrightarrow \mathcal{E}_m$
\begin{equation}
\begin{array}{l}
\tilde{a}_m|n\rangle =\left\{ \begin{array}{ll}
0 & {\rm if} \ \ n=0\\
\sqrt{\tilde{e}_n}\,{\rm e}^{{\rm i}\gamma (e_n-e_{n-1})} |n-1\rangle & {\rm if} \ \ n>0
\end{array} \right. \nonumber \\[5mm]
\tilde{a}_m^+|n\rangle \!=\! \left\{ \begin{array}{ll}
\sqrt{\tilde{e}_{n+1}}\, {\rm e}^{-{\rm i}\gamma (e_{n+1}-e_n)} |n+1\rangle & {\rm if}\ \  n\!<\!\mathcal{L}\\
0 & {\rm if} \ \ n\!=\!\mathcal{L}
\end{array}\right. 
\end{array}
\end{equation}
where $\tilde{e}_n=n(\mathcal{L}-n+1)$ and 
\begin{equation}
e_n=\lambda _{m+n}-\lambda _m=-n(n+2m+\alpha -1).
\end{equation}

These operators satisfy the relations
\begin{equation}\begin{array}{l}
[\tilde{a}_m^+,\tilde{a}_m]=2\, \tilde{R}_m\\[2mm]
[\tilde{R}_m,\tilde{a}_m^+]=\tilde{a}_m^+\\[2mm]
[\tilde{R}_m,\tilde{a}_m]=-\tilde{a}_m
\end{array}
\end{equation}
where $\tilde{R}_m$ is the operator 
$\tilde{R}_m=N-\frac{ \mathcal{L} }{2}$
defined by using the `number operator'
\begin{equation}
N:\mathcal{E}_m\longrightarrow \mathcal{E}_m\qquad N|n\rangle =n|n\rangle  .
\end{equation}
Therefore, the Lie algebra generated by $\tilde{a}_m^+$ and $\tilde{a}_m$ is isomorphic to $su(2)$.
The functions $|n\rangle $ are eigenfunctions of the operator
$H=H_m-\lambda _m$
\begin{equation}
H|n\rangle =e_n|n\rangle .
\end{equation}
\mbox{}\\[5mm]
\noindent{\large \bf 5.2 \ Coherent states}\\
By following [16] and the analogy with the case $\sigma (s)\in \{ 1,\, s,\, 1-s^2\}$,
we consider for each $z\in \mathbb{C}$ the function
\begin{equation}
 |z, \gamma \rangle=\sum_{n=0}^\mathcal{L} \frac{z^n\, 
{\rm e}^{-{\rm i}\gamma e_n}}{\sqrt{\tilde{\varepsilon }_n}}|n\rangle 
\end{equation}
where
\begin{equation}
 \tilde{\varepsilon }_n=\left\{ \begin{array}{lll}
1 & {\rm if} & n=0\\
\tilde{e}_1\tilde{e}_2...\tilde{e}_n & {\rm if } & n>0\, .
\end{array} \right. 
\end{equation}
The system  $\{ \ |z, \gamma \rangle \ |\ z\in \mathbb{C}\}$ is an overcomplete 
system of functions in the finite dimensional space $\mathcal{E}_m$ with 
\begin{equation}
\langle z, \gamma |z, \gamma \rangle =\sum_{n=0}^\mathcal{L}\frac{|z|^{2n}}{\tilde{\varepsilon }_n}
\end{equation}
and the property (temporal stability)
\begin{equation}
 {\rm e}^{-{\rm i}tH}|z, \gamma \rangle =|z, \gamma +t\rangle .
\end{equation}
It can be regarded as a system of coherent states in $\mathcal{E}_m$.\\[1.2cm]
\noindent{\Large \bf 7 \ Concluding remarks}\\ 
The associated hypergeometric-type functions can be studied together
in a unified formalism, and are directly related to the bound-state 
eigenfunctions of some important Schr\"odinger equations 
(P\"oschl-Teller, Morse, Scarf, etc.).

It is useful to obtain fundamental versions (at the level of
associated special functions) for some methods and formulae from 
quantum mechanics because in this way one can extend results known in
particular cases to other quantum systems. A large number of formulae 
occurring in various applications of quantum mechanics follow from a very small
number of fundamental mathematical results.\\[5mm]
\noindent{\small {\it Acknowledgment:}\\ 
This work was supported by the grant CERES 4-192.}\\[5mm]

\noindent{\it References:}
\begin{itemize}
\item[ [1]]
J-P. Antoine, J-P. Gazeau, P. Monceau, J. R. Klauder, and K. A. Penson,
Temporally stable coherent states for infinite well and P\"oschl-Teller potentials,
{\it J. Math. Phys.}, 42, 2001, pp.~2349--2387.

\item[ [2]]
A.~O. Barut and L. Girardello, New ''cohe-rent'' states associated
with non-compact groups, {\it Commun. Math. Phys.}, 21, 1971, pp.~41--55.

\item[ [3]]
V. Bargman, On a Hilbert space of analytic functions and an associated 
integral transform, {\it Commun.Pure Appl. Math.}, 14, 1961, p. 187.

\item[ [4]]
H. Bateman, {\it Integral transformations}, vol. I,  
Erdelyi, ed., McGraw-Hill, New York, 1954, p. 349.

\item[ [5]]
F. Cooper, A. Khare, and U. Sukhatme, Supersymmetry and quantum 
mechanics,  {\it Phys. Rep.}, 251, 1995, pp.~267--385.

\item[ [6]]
N. Cotfas, Shape invariance, raising and lowering operators in
hypergeometric type equations, {\it J. Phys. A: Math. Gen.}, 35, 2002, pp.~9355--9365.

\item[ [7]]
N. Cotfas, Systems of orthogonal polynomials defined by hypergeometric
type equations with application to quantum mechanics, 
{\it Cent. Eur. J. Phys.}, 2, 2004, pp.~456--466.

\item[ [8]]
M. Daoud and V. Hussin, General sets of coherent states and the 
Jaynes-Cummings model, {\it J. Phys. A: Math. Gen.}, 35, 2002, pp.~7381--7402.

\item[[9]]
J-P. Gazeau and J. R. Klauder, Coherent states for systems with discrete 
and continous spectrum, {\it J. Phys. A: Math. Gen.}, 32, 1999, pp.~123--132.

\item[ [10]]
M.~N.~H. Hounkonnou and K. Sodoga,
Generalized coherent states for associated hypergeometric-type functions,
{\it J. Phys. A: Math. Gen.}, 38, 2005, pp.~7851--7862.

\item[ [11]]
L. Infeld  and T.~E. Hull, The factorization method,
{\it Rev. Mod. Phys.},  23, 1951, pp.~21--68.

\item[[12]]
M.~A. Jafarizadeh and H. Fakhri,   
Parasupersymmetry and shape invariance in differential equations of
mathematical physics and quantum mechanics,
{\it Ann. Phys. (N. Y.)}, 262, 1998, pp.~260--276.

\item[[13]]
A.~H. El Kinani and M. Daoud, Generalized coherent and intelligent states for exact solvable 
quantum systems, {\it J. Math. Phys.}, 43, 2002, pp.~714--733.

\item[[14]]
M. Lorente, Raising and lowering operators, factorization and 
differential/difference operators of hypergeometric type,
{\it J. Phys.: Math. Gen.}, 34, 2001, pp.~569--588.

\item[[15]]
A.~F. Nikiforov, S.~K. Suslov, and V.~B. Uvarov, 
{\it Classical Orthogonal Polynomials of a Discrete Variable}, Springer,
Berlin, 1991.

\item[[16]]
B. Roy and P. Roy, Gazeau-Klauder coherent states for the Morse potential 
and some of its properties, {\it Phys. Lett. A}, 296, 2002, pp.~187--191.
\end{itemize}
\end{document}